\documentclass[12pt]{iopart}
\newcommand{\bal}{\mbox{\boldmath $\alpha$}}
\newcommand{\bsi}{\mbox{\boldmath $\sigma$}}

\usepackage{graphics}
\usepackage{cite}

\begin{document}
\title[Dirac Oscillator]
{The Dirac Oscillator. A relativistic version of the Jaynes--Cummings
model}
\author{P Rozmej\dag\footnote[4]{E-mail: rozmej@tytan.umcs.lublin.pl
and arvieu@in2p3.fr}  and R Arvieu\ddag} 

\address{\dag Instytut Fizyki,  Uniwersytet MCS,  20-031 Lublin,  Poland}
\address{\ddag Institut des Sciences Nucl\'eaires,  F-38026 Grenoble,  France}
\begin{abstract} 
 The dynamics of wave packets in a relativistic Dirac oscillator
is compared to that of the Jaynes-Cummings model. The strong spin-orbit
coupling of the Dirac oscillator produces the entanglement of the spin
with the orbital motion similar to what is observed in the model of
quantum optics. The collapses and revivals of the spin which result extend
to a relativistic theory our previous findings on nonrelativistic
oscillator where they were known under the name of {\em spin-orbit
pendulum}. There are important relativistic effects (lack of
periodicity, zitterbewegung, negative energy states). Many of them disappear
after a Foldy-Wouthuysen transformation.
\end{abstract}
%\submitted 
\section{Introduction}\label{intro}
  
 This paper is an essay to mix together different popular models of
quantum theory which have been developed separately and on different
purposes and to present the Dirac oscillator as a relativistic version of
the Jaynes--Cummings (JC) model with, in addition to the regular properties
of this model, some interesting new ones related to the relativistic
description.  
 The Dirac oscillator, (DO) first introduced by Ito et al. 
 \cite{ito},  %[1] 
 was later shown by Cook \cite{cook} %[2] 
to present unusual accidental degeneracies in its
spectrum which were discussed from a supersymmetric viewpoint by 
Ui et al \cite{ui}. %[3]. 
It was refreshed later by Moshinsky and Szczepaniak \cite{szczep} %[4] 
and its symmetry Lie algebra was explicited by 
Quesne and Moshinsky \cite{quesne}. %[5]. 
Moreno and Zentella \cite{moreno} %[6] 
showed also that an exact Foldy-Wouthuysen (FW) transformation
could be performed. More recently Nogami and Toyama \cite{nogami} %[7] 
and Toyama et al \cite{toyama} %[8] 
have studied the behaviour of wave packets (WP) of the DO in the Dirac
representation and in the FW representation in (1+1) dimensions. The aim of
these authors was to study WP which could possibly be coherent. This
reduction of the dimension was brought as an attempt to get rid of spin
effects and to concentrate on the relativistic effects. 
 
 Our aim is to extend the work of \cite{nogami,toyama} %[7-8] 
 and to consider the full (3+1)
dimensions and to show that we can there make an interesting new
connection with the JC model as we will explain shortly. 

 The degeneracies of the eigenvalues of the DO are due to a spin orbit
potential which is unusually large. In previous papers
\cite{arv94,arv95,rozm96} %[9-11] 
we have analyzed the time dependent behaviour of WP 
in a nonrelativistic harmonic
oscillator potential with a constant spin orbit potential. We have shown
that the behaviour of the spin shares a strong analogy with the
observations made on the population of a two level atom in a cavity where
it can make a two photons exchange \cite{knight}. %[12]. 
 
 For the relativistic DO the mechanism of collapses and revivals well
known in the JC model is then expected to take place with some
differences. [see \cite{shore} %[13] 
for a general review on the JC model and for a full list of references]. 
Since the energy spectrum is far from being
linear the exact periodicity of the nonrelativistic oscillator is lost and
the system evolves more on the lines of the regular JC model i. e.  without
exact recurrences. In addition to that, the spin motion should exhibit the
famous zitterbewegung, this trembling motion should also be seen in the
motion of the density of the wave as shown in \cite{nogami,toyama}. %[7-8]. 
This effect should disappear in the FW representation.
   
   Klein paradox is a second typical behaviour type 
presented by a relativistic WP scattered by a barrier. 
There it is generally said that it is a point % place 
in which Dirac equation cannot be interpreted as describing a single
particle. This effect is more conveniently explained in the frame of hole
theory. Studying the behaviour of a WP of the DO in a highly relativistic
regime we have found that the WP contains a component counter-rotating
that can indeed be interpreted with hole theory and which disappears in
the FW representation.    

\section{Summary of results on the DO} \label{se2} 
  
  In the following we will use the notations of \cite{szczep} %[4] 
and of the book \cite{moshinsky}. %[14]. 
The time dependent Dirac equation is written as:
\begin{equation}\label{l1} %(1)
i \hbar\, \frac{\partial \Psi}{\partial t} = H_D\, \Psi =
 c \,[ \,\bal \cdot (\bi{p} -i m \omega\bi{r}\beta)+ mc\beta\,] \,\Psi  \: .
\end{equation}
 The components $\Psi_1$ and $\Psi_2$ of a spinor of energy $E$ 
\begin{equation}\label{l2} %(2)
 \Psi = \left( \begin{array}{c} \Psi_1 \\ \Psi_2 \end{array} \right)  
\end{equation}
 obey the equations
\numparts
\begin{eqnarray}\label{l3} %(3)
 ( E^2 - m^2 \, c^4 )\, \Psi_1 & = & [c^2(\bi{p}^2+m^2\omega^2 \bi{r}^2)  
 -3\hbar\omega mc^2 - \frac{4 mc^2\omega}{\hbar}  
 (\bi{L}\cdot\bi{S}) ]\,\Psi_1 \\
 ( E^2 - m^2 \, c^4 )\, \Psi_2 & = & [c^2(\bi{p}^2+m^2\omega^2 \bi{r}^2)  
 +3\hbar\omega mc^2 + \frac{4 mc^2\omega}{\hbar} 
 (\bi{L}\cdot\bi{S}) ]\,\Psi_2    \, .
\end{eqnarray}
\endnumparts
These components are thus the eigenstates of a spherical HO with a spin
orbit coupling term respectively $\pm\,2\omega/\hbar$. 
These large coupling
strengths are responsible of the unusual degeneracies of the levels. 
The spectrum depends on a single parameter $r$ defined as 
\begin{equation}\label{l4} %(4)
            r =\frac{\hbar\omega}{mc^2}       \: .  
\end{equation}

  Spectrum and degeneracies and the building of $\Psi_1$ are well described in
\cite{szczep} and \cite{quesne}. %[4] and [5]. 
For an eigenstate of energy $E_{nlj}$ $\Psi_1$ can also be labelled by
$n$, the usual total number of quantas of the 3D oscillator,  by the orbital
and total angular momentum $l$ and $j$ and by the component $m$ of $j_z$. 
In terms of $r$ 
\begin{equation}\label{l5} %(5)
          E_{nlj} = mc^2\,\sqrt{r A+1}   \:,  
\end{equation}
where $A$ is defined as
\numparts
\begin{eqnarray}\label{l6} %(6)
  A&=&2(n-j)+1  \quad\quad {\rm if} \quad\quad l=j-\case{1}{2}    \\
  A&=&2(n+j)+3  \quad\quad {\rm if} \quad\quad l=j+\case{1}{2}   \: .
\end{eqnarray}
\endnumparts
   Thus the states which obey (6a) have an infinite degeneracy. Among them
those with $n=l$ have the lowest value $E=mc^2$. The states which obey [6b]
have on the contrary a finite degeneracy. 
 
    For the eigenvalues of eq.~(3b) one should take care of the opposite
sign of the spin orbit potential and of the different sign of the 
constant term. The eigenvalue is written in terms of $n' \, l'$ and $j'$ as
\begin{equation}\label{l7} %(7)
         E_{n'l'j'}=mc^2\, \sqrt{r A'+1}    \:,    
\end{equation}
   with $A'$ given by 
\numparts
\begin{eqnarray}\label{l8} %(8)
   A'& = & 2(n'-j')+3	\quad\quad {\rm if} \quad\quad l'=j'-\case{1}{2}   \\
   A'& = & 2(n'+j')+5	\quad\quad {\rm if} \quad\quad l'=j'+\case{1}{2}   \: .
\end{eqnarray}
\endnumparts

      The need to make $E$ and $E'$ equal with different $n$ and $n'$, 
      $l$ and $l'$ is 
fulfilled if one notices that $\Psi_2$ is connected to $\Psi_1$ by
\begin{equation}\label{l9} %(9)
 |\Psi_2\rangle = -i\, \frac{mc^2}{E+mc^2}\,\sqrt{2r}(\bsi \cdot \bi{a})
 \,|\Psi_1\rangle \: .
\end{equation}
The operator which couples $\Psi_1$ and $\Psi_2$ above is expressed as 
the scalar product of the spin operator 
$\bsi$ with a vector annihilation operator $\bi{a}$ defined by
\begin{equation}\label{l10} %(10)
     \bi{a}= \frac{1}{\sqrt{2}}\left[ \frac{\bi{r}}{\sqrt{\hbar/m\omega}}
     + i\, \frac{\bi{p}}{\sqrt{\hbar m\omega}} \right]   \: .           
\end{equation}
 In order to satisfy eqs.~(\ref{l3}) and (\ref{l9}) it is necessary that
\numparts
\begin{eqnarray}\label{l11} %(11)
    j'&=& j  \quad\quad   n'=n-1    \\ % (11a)
   l'&=&l+1 \quad\quad {\rm if} \quad\quad  l=j-\case{1}{2}    \\ % (11b)
   l'&=&l-1 \quad\quad {\rm if} \quad\quad  l=j+\case{1}{2}    \: .    % (11c)
\end{eqnarray}
\endnumparts
  These conditions imply that for the ground state of the equation
satisfied by $\Psi_1$ one has $\Psi_2 = 0$. 
   This corresponds to the fact that the DO is a supersymmetric potential
for which $\Psi_1$ and $\Psi_2$ have the same spectrum apart from  the 
absence in the spectrum of $\Psi_2$ of the ground state eigenvalue of $\Psi_1$. 

 Beside eq.~(\ref{l9}) we have 
\begin{equation}\label{l12} %(12) 
 |\Psi_1\rangle = i\, \frac{mc^2}{E-mc^2}\,\sqrt{2r}(\bsi \cdot \bi{a}\dag)
 \,|\Psi_2\rangle \: .
\end{equation}
  Using $|\Psi_1\rangle=|nljm\rangle$ and inserting (\ref{l9}) into 
  (\ref{l12}) one obtains
\begin{equation}\label{l13} %(13) 
  E^2-m^2c^4 = (mc^2)^2 \, 2r \, |\langle n'l'jm |(\bsi \cdot \bi{a})|
  nljm \rangle|^2  \: ,                      
\end{equation}
where $|n'l'jm\rangle$ is a normalized harmonic oscillator state which has its
quantum numbers defined by (\ref{l11}). 
From (\ref{l9}) and (\ref{l13}) and using the phase 
conventions as in the reference book \cite{moshinsky} % [14] 
one obtains
\begin{equation}\label{l14} %(14) 
   \langle n'l'jm|\Psi_2\rangle = sgn \sqrt{\frac{E-mc^2}{E+mc^2}}  \: .     
\end{equation}
The complex phase called $sgn$ is defined by
\numparts
\begin{eqnarray}\label{l15} %(15)
       sgn&=&-i \quad\quad {\rm if} \quad\quad l'=l+1=j+\case{1}{2}  \\ %   (15a)
       sgn&=&+i \quad\quad {\rm if} \quad\quad l'=l-1=j-\case{1}{2}  \: . % (15b)
\end{eqnarray}
\endnumparts

  A normalized spinor with positive energy $E=+E_p$ can now be expressed as
\begin{equation}\label{l16} %(16) 
\Psi_{+}(t) = \left[ \begin{array}{r}
\sqrt{\frac{E_p+mc^2}{2E_p}}\, |nljm\rangle \\
sgn \sqrt{\frac{E_p-mc^2}{2E_p}}\, |n'l'jm\rangle \end{array}
\right] \exp \left(\frac{-iE_p t}{\hbar}\right) \: . % (16)
\end{equation}
  In a similar manner a spinor with negative energy $E=-E_p$ is written as        
\begin{equation}\label{l17} %(17) 
\Psi_{-}(t) = \left[ \begin{array}{r}
\sqrt{\frac{E_p-mc^2}{2E_p}}\, |nljm\rangle \\
-sgn \sqrt{\frac{E_p+mc^2}{2E_p}}\, |n'l'jm\rangle \end{array}
\right] \exp \left(\frac{iE_p t}{\hbar}\right) \: . % (17)
\end{equation}
  It is interesting to note that the relative weights of the large and
small components are formally expressed in terms of $E_p$ exactly in the same
manner as in the (1+1) model of \cite{nogami,toyama}. %[7-8]. 
The energies are however given by eq.~(\ref{l5})  with conditions (6a-6b). 

\section{Study of a circular wave packet. Theory} \label{se3}
  
\subsection{Definition of the WP} \label{ss1}

 In the following we will study the time evolution of a circular WP of a
special kind in the DO. For $t=0$ we assume that the WP has a gaussian shape
with an average position $\bi{r}_0$ and an average momentum $\bi{p}_0$ 
such that the WP
moves upon a circular trajectory if it is left free of spin and in a
nonrelativistic HO. The WP is assumed to be initially an eigenstate
of spin with an arbitrary direction defined by two complex numbers $\alpha$
and $\beta$. Let the normalized WP be
\begin{equation}\label{l18} %(18) 
\Psi(\bi{r},0) = \frac{1}{(2\pi)^{3/4}\,\sigma^{3/2}}
\, \exp \left[ -\frac{(\bi{r}-\bi{r}_0)^2}{2\sigma^2} 
 + i \, \frac{\bi{p}_0\cdot\bi{r}}{\hbar}\right]
 \left( \begin{array}{r} \alpha \\ \beta\\ 0 \\ 0 \end{array} \right)
\end{equation}
It is simpler to choose the axis of coordinates such that
\begin{equation}\label{l19} %(19) 
  \bi{r}_0=\bi{x}\,x_0  \: . 
\end{equation}
The centroid $x_0$ is expressed in units of the natural width of the HO
and of a parameter called $N$ by
\begin{equation}\label{l20} %(20) 
 x_0 = \sqrt{N}\, \sigma = \sqrt{N}\,\sqrt{\hbar/(m\omega)} \: , 
\end{equation}
while the average momentum is taken as  
\begin{equation}\label{l21} %(21) 
 \bi{p}_0 = \bi{y}\, p_0 = \bi{y}\,\hbar\,\sqrt{N}/\sigma \: .   
\end{equation}
In (\ref{l19}) and (\ref{l21}) $\bi{x}$ and $\bi{y}$ denote the 
appropriate unit vectors.
  
The average angular momentum is then 
\begin{equation}\label{l22} %(22) 
 \langle L_z \rangle = x_0\,p_0 = N \, \hbar    \: .
\end{equation}
The partial wave expansion of the WP involves only waves 
for which $m=l$ and
for which the total number of quantas of the oscillator is also $l$. 
It is given by 
\begin{equation}\label{l23} %(23) 
  |\Psi(0)\rangle = \sum_{l=0}^{\infty} \, \lambda_l \,|n=l\,l\,m_l=l\rangle      
 \left( \begin{array}{r} \alpha \\ \beta\\ 0 \\ 0 \end{array} \right) \: .
\end{equation}
         The weights $\lambda_l$ are given by
\begin{equation}\label{l24} %(24)
	\lambda_l = (-1)^l \, \exp(-N/2)\, \frac{N^{l/2}}{\sqrt{l!}}  \: .
\end{equation}

 In other words (\ref{l18}) is a coherent state of the HO. We have studied in
\cite{arv94,arv95,rozm96} % [9-11] 
its time evolution assuming that the hamiltonian is a non
relativistic HO with a constant spin-orbit potential which depends upon a
parameter called $\kappa$ as
\begin{equation}\label{l25} %(25)
                  V_{s.o.} = \kappa\,(\bi{L}\cdot \bsi) \:. % (25)
\end{equation}
 Associated to the DO there is a nonrelativistic hamiltonian that we will
define as the operator in the right side of (3a) divided by $2mc^2$ and
which thus presents the same degeneracies as those given by eq. (6a-b). 
  
 \subsection{Partial waves with spin up} \label{ss2}
  
   Let us isolate in the WP (\ref{l18}) a partial wave $l$ with spin up. 
A partial wave has two partners with different total angular momentum called
$j_+=l+\case{1}{2}$ and $j_-=l-\case{1}{2}$ with respective energies 
$E_+$ and $E_-$. $E_+$ is given for
all the $l$ present in (\ref{l23}) by
\numparts
 \begin{eqnarray}\label{l26} %(26)
  E_+ & = & mc^2\,  \quad\quad   {\rm in~the~relativistic~theory} \\ %  (26a)
  E_+ & = & 0  \quad\quad\quad {\rm ~in~the~nonrelativistic~one} \: . %(26b)  
 \end{eqnarray} 
\endnumparts  
A partial wave with spin up has $j=j_+$ and is an eigenstate of the
hamiltonian with energy $E_+$. It leads trivially to the phase
$\exp(-iE_+ t) =\exp(-i\omega_0 t)$ in case (26a) with $\omega_0=mc^2/\hbar$.
 
 \subsection{Partial waves with spin down} \label{ss3}  
 
   This component is coupled to the two partners and we need therefore $E_-$
which is given by
 \numparts
 \begin{eqnarray}\label{l27} %(27)
  E_- & = & mc^2\sqrt{2r(2l+1)+1} \quad\quad\quad\quad\quad 
  {\rm in~the~relativistic~theory} \\ % (27a)
  E_- & = & mc^2\,r(2l+1)=\hbar\omega (2l+1) \quad\quad
  {\rm in~the~nonrelativistic~one} \: .  %(27b) 
 \end{eqnarray}
\endnumparts  

{\bf i)} Let us consider first a two component spinor in the nonrelativistic
theory 
\begin{equation}\label{l28} %(28)
 \left( \begin{array}{c} 0  \\ |l\,l\,l\rangle \end{array} \right)
 = \frac{1}{\sqrt{2l+1}}\, |l\,j_+\,m_j=j_+-\case{1}{2}\rangle 
 + \frac{2l}{\sqrt{2l+1}}\, |l\,j_-\,m_j=j_-\rangle \: . % (28)
 \end{equation}
After a time $t$ the spinor is given by
\begin{equation}\label{l29} %(29)
 \left( \begin{array}{c} 0  \\ |l\,l\,l\rangle \end{array} \right)_t
 = \frac{1}{2l+1}\, \left( \begin{array}{c} 
 \sqrt{2l}\,(e^{-iE_+t/\hbar}-e^{-iE_-t/\hbar}) \,|l\,l\,m_l=l-1\rangle \\
 (e^{-iE_+t/\hbar} +2l\,e^{-iE_-t/\hbar}) \,|l\,l\,m_l=l\rangle \end{array}
 \right)  \: . %  (29)
\end{equation}
 
  One sees that the average components of $\bsi$ are given by    
 \numparts
 \begin{eqnarray}\label{l30} %(30)
  \langle\sigma_x\rangle&=& 0  \quad\quad \langle \sigma_y\rangle =0  \\ % (30a)
  \langle\sigma_z\rangle&=&-1+\frac{16l}{(2l+1)^2}\,\sin^2[(2l+1)\,\omega t]
  \: . % (30b)
 \end{eqnarray}
\endnumparts  
  At times  
\begin{equation}\label{l31} %(31)
               t= \pi/[2\omega(2l+1)]+n\pi  % (31)
 \end{equation}
 the average of the spin is zero and the spin has a maximum of
entanglement with the orbital motion. 
  The most important component in the spinor (\ref{l29}) is for large enough
values of $l$ the spinor
\begin{equation}\label{l32} %(32)
 \left( \begin{array}{c} 0  \\  \frac{2l}{2l+1}\, e^{-i\omega (2l+1)t} \,
 |l\,l\,l\rangle \end{array} \right)  \: . % (32)
\end{equation}
                          
 The time dependent part of the phase $\exp(-i2\omega t)$ in this spinor can be
incorporated with the angular phase $\exp(il\phi)$ of the spherical
harmonic. Therefore the main effect for this partial wave with spin down is
that it rotates  around $Oz$ with angular velocity $2\omega$.  
  Thus we have shown that a partial wave with arbitrary $\alpha$ and $\beta$
contains a part which is at rest and a second part which rotates around $Oz$
with angular velocity $2\omega$. This is a particular case of our previous
findings \cite{rozm96}: % [11]:
in the general case with arbitrary $\kappa$ the two waves with
opposite spin move around $Oz$, the spin up part with angular velocity
$\omega-\omega_{ls}$ and the spin down one with $\omega+\omega_{ls}$. 
In the DO we have simply $\omega_{ls}=\omega$. 
  
{\bf ii)} The corresponding relativistic spinor is written simply by adding two
components which are zero initially.
\begin{equation}\label{l33} %(33)
 \left( \begin{array}{c} 0  \\|l\,l\,l\rangle \\0 \\0 \end{array} \right)  
 = \frac{1}{\sqrt{2l+1}}
 \left( \begin{array}{c} |l\,j_+\,j_+-1\rangle \\0 \\0 \end{array} \right)
 + \sqrt{\frac{2l}{2l+1}}
 \left( \begin{array}{c} |l\,j_-\,j_-\rangle \\0 \\0 \end{array} \right)   
 \: . % (33)
\end{equation}
 The first spinor gets the phase $\exp(-i\omega_0 t)$ while the second spinor
requires an expansion in terms of spinors with positive and negative
energy as analyzed in \cite{nogami,toyama}. % [7-8]. 
The coupling is expressed by using eq.~(\ref{l16}) and
(\ref{l17}) in terms of a coefficient called $a_l$ defined as
\begin{equation}\label{l34} %(34)
 a_l = \frac{\sqrt{E_-^2-m^2 c^4}}{E_-} = \sqrt{\frac{2r(2l+1)}{1+2r(2l+1)}}
  = \sqrt{1-\left( \frac{\omega_0}{\omega_l}\right)^2}  \: , % (34) 
\end{equation}
where we have defined $\omega_l$ by
\begin{equation}\label{l34a} %(34a)
 \omega_l = \omega_0\,\sqrt{1+2r(2l+1)}\: .
\end{equation}
 The second term of (\ref{l33}) is written at time $t$ as
\begin{eqnarray}\label{l35} %(35)
\left( \begin{array}{c} |l\,j_-\,j_-\rangle \\0 \\0 \end{array} \right)_t
 & = &   \left( \begin{array}{c} 
    (\cos \omega_l t -i\frac{\omega_0}{\omega_l}\,\sin \omega_l t) \,
    |l\,j_-\,j_-\rangle  \\
    sgn \, a_l \,\sin \omega_l t \, |l-1\,j_-\,j_-\rangle \end{array} \right)\\
 & = &  \left( \begin{array}{c}
    - \frac{1}{\sqrt{2l+1}}(\cos \omega_l t -i\frac{\omega_0}{\omega_l}\,
    \sin \omega_l t) \, |l\,l\,l-1\rangle  \\
     \sqrt{\frac{2l}{2l+1}}(\cos \omega_l t -i\frac{\omega_0}{\omega_l}\,
     \sin \omega_l t) \, |l\,l\,l\rangle  \\
     sgn \, a_l\, \sin \omega_l t \, |l-1\,l-1\,l-1\rangle  \\
     0  \end{array} \right) \:. 
  \label{l36}  % (36)
\end{eqnarray}
Finally reassembling the four parts of the spinor (\ref{l33}) one writes it as
\begin{equation}\label{l37} %(37)
\fl 
 \left( \begin{array}{c} 0  \\|l\,l\,l\rangle \\0 \\0 \end{array} \right)_t  
 = \frac{1}{2l+1} \left( \begin{array}{c} {}
[ \sqrt{2l}\,[ e^{-i\omega_0 t}-(\cos \omega_l t -i\frac{\omega_0}{\omega_l}\,
 \sin \omega_l t) ] \, |l\,l\,l-1 \rangle  \\
 {} [e^{-i\omega_0 t} + 2l
 (\cos \omega_l t -i\frac{\omega_0}{\omega_l}\,\sin \omega_l t)] 
 \, |l\,l\,l \rangle  \\
 \sqrt{2l(2l+1)} \, sgn \, a_l \, \sin \omega_l t \,|l-1\,l-1\,l-1\rangle \\ 0
 \end{array} \right) \: .
\end{equation}
Comparing (\ref{l37})  to (\ref{l29}) we see that time evolution 
creates a small
component in the relativistic spinor which is zero initially while the
effect in the large components is through the following replacement
\begin{eqnarray}\label{l38} %(38)
 e^{-iE_- t/\hbar} 
 &\rightarrow & (\cos \omega_l t -i\frac{\omega_0}{\omega_l}\, \sin \omega_l t) \\
 &=& \case{1}{2}\, e^{-i\omega_l t}[1+ \frac{\omega_0}{\omega_l}]
  +  \case{1}{2}\, e^{i\omega_l t}[1- \frac{\omega_0}{\omega_l}] \: .
  \label{l39}  % (39)
\end{eqnarray}

  With similar arguments as those which lead us to eq.~(\ref{l32}) 
we obtain in the relativistic case two waves which rotate in opposite 
sense with angular velocity $2\omega$.
To get this conclusion we must linearize $\omega_l$ by writing
\begin{equation}\label{l40} %(40)
  \omega_l=\omega_0\,[1+r(2l+1)]=\omega_0+\omega (2l+1)  \: .    
\end{equation}
 The part with positive frequency is now weighted by 
 $\case{1}{2}(1+\omega_0/\omega_l)$ 
and the new part which rotates in opposite sense by
$\case{1}{2}(1-\omega_0/\omega_l)$.
The third component of the spinor (\ref{l37}) contains also
two waves with opposite sense of rotation and amplitude $a_l/2$. 
   Thus a relativistic WP with $\alpha=\beta=1/\sqrt{2}$ 
dissociates into three parts instead of two as in the 
nonrelativistic evolution: there is a part
with spin up which does not move essentially, 
while the part with spin down
is divided in two waves moving in opposite directions. 
    
 \subsection{Spin averages} \label{ss4}
  
  We give below the expectation values of the operator $\bsi$ for the
circular WP (\ref{l23}) assuming $\alpha=\beta=1/\sqrt{2}$
\begin{equation}\label{l41} %(41)
\fl \langle\sigma_x\rangle = \sum_l \, |\lambda_l|^2 \, \frac{1}{2l+1}
 \left[ 1+l(1+\frac{\omega_0}{\omega_l}) \cos (\omega_l-\omega_0)t
         +l(1-\frac{\omega_0}{\omega_l}) \cos (\omega_l+\omega_0)t 
 \right]  \: ,    
\end{equation}
\begin{equation}\label{l42} %(42)
\fl \langle\sigma_y\rangle = \sum_l \, |\lambda_l|^2 \, \frac{1}{2l+1}
 \left[ (1+\frac{\omega_0}{\omega_l}) \sin (\omega_l-\omega_0)t
       +(1-\frac{\omega_0}{\omega_l}) \sin (\omega_l+\omega_0)t
 \right]  \: ,    
\end{equation}
\begin{eqnarray}\label{l43} %(43)
 \langle\sigma_z\rangle = & \sum_l \, |\lambda_l|^2 \, \frac{1}{2l+1}
 & \left \{  
 \frac{1}{2} + \frac{4l-1}{2(2l+1)^2}- \frac{\omega_0^2}{\omega_l^2}
 \frac{2l^2}{2(2l+1)^2} \right. \nonumber \\ & & 
-\frac{2l}{(2l+1)^2}(1+\frac{\omega_0}{\omega_l})\cos(\omega_0-\omega_l)t
 \\& &
-\frac{2l}{(2l+1)^2}(1-\frac{\omega_0}{\omega_l})\cos(\omega_0+\omega_l)t
 \nonumber \\& &  \left.
 -\frac{2l^2}{(2l+1)^2}(1-\frac{\omega_0^2}{\omega_l^2})
 \cos 2\omega_l t  \right \} \nonumber  \: .    
\end{eqnarray}
   These formulas extend to the relativistic DO those we have already
discussed in \cite{arv94}. %[9].
Because of the conservation of the total angular momentum
there is no interference between the various $l$ for the spin averages.
Each partial wave depends on time because the energies of the 
spin orbit partners are different. 
In the nonrelativistic case the time factors depend
only on the differences $\omega_l-\omega_0=(2l+1)\omega$.
Therefore all these averages have period $2\pi/\omega$.
After a time called the collapse time
$\tau_c=\pi/(2\sqrt{2N})$ 
all the phases coming from all the partial waves
are equally distributed and assuming high values of $N$ all these averages
are zero, i.e.\ there is a collapse of the spin!
The average orbital angular
momentum gets correspondingly an increase in order to preserve the average
total angular momentum. This exchange was called 
{\em the spin orbit pendulum}
since it occurs exactly periodically. In addition we have also shown that
for a time equal to $\pi/\omega$ the average spin is for high $N$ 
opposite to
its initial value with the same coherence time around this revival. This
behaviour make the spin orbit pendulum analogous to the JC model with the
frequencies $(2l+1)\omega$ playing the role of the Rabi frequencies.

   For the DO we obtain essentially the same behaviour. However the
periodicity is totally broken for high and even for low values of $r$
because of the terms involving $\omega_0+\omega_l$. 
This combination introduces
in the spin motion high frequencies affecting each component with a
different weight. The effect of this modulation is the well known
{\em zitterbewegung} that can thus be seen in the DO on the observable of
spin. Note that (\ref{l41}) and (\ref{l42})  contain similar terms 
and similar weights. $\langle\sigma_x\rangle$ and
$\langle\sigma_y\rangle$
will present then a similar time behaviour.
The $Oz$ component is however different since there is in (\ref{l43}) an
extra term with frequency $2\omega_l$. 
This term will produce two effects: an extra high frequency modulation 
and a partial revival at a time about half of the revival of the spin.
This new revival of a purely relativistic
origin concerns only the $z$ component which will oscillate rapidly while
the other components are zero. 

\section{Circular WP. Numerical calculations}

  In order to exhibit the various effects discussed in the preceding
section we have chosen the value $N=20$ which provides a WP well
concentrated in configuration space with an interesting spread of its
partial waves. This value was also used thoroughly in our previous
papers on the nonrelativistic oscillator \cite{arv94,arv95,rozm96}. % [9-11].
For simplicity we use the units $\hbar=m=c=1$. Therefore our time units
used in presented figures are proportional to $\omega^{-1}=r^{-1}$.
For nonrelativistic case the period of the motion is then $T=2\pi/\omega
=2\pi/r$. 

\subsection{Spin averages in the DIRAC representation} \label{s4s1}
   For the very low values of $r$, like $r=0.001$ in Fig.~\ref{r001}, %1, 
we are very near the nonrelativistic limit. 
One observes then a collapse of each
component of the spin during an interval of time $\tau_c$ and a revival of the
spin which has lost its periodicity because of the use of the relativistic
energies. The first revival of $\langle\sigma_x\rangle$ and 
$\langle\sigma_y\rangle$ occurs around time
$\pi/\omega$. In the nonrelativistic case there was \cite{arv94} %[9] 
a revival of $\langle\sigma_x\rangle$
with a change of sign. In the present case the spin rotates rapidly in the
plane $xOy$ and its maximum value is with the same sign as initially. The
relativistic effects produce a slow decrease in the amplitude of the
revivals. There is already a quite sensible difference in the behaviour of
$\langle\sigma_z\rangle$ with time. 
This component fluctuates much more rapidly because it
is richer in frequencies than the other two. It exhibits also a small
increase at a time about half of the recurrence time due to these higher
components. 

 Each of these effects becomes more pronounced when the parameter $r$ is
given higher values. Figure~\ref{r025} %2 
is for $r=0.025$, Fig.~\ref{r5} for $r=0.5$. The components of
the spin in the $xOy$ plane oscillate much longer around each recurrence
with a small period and the amplitudes of these recurrences decay. Again
the behaviour of $\langle\sigma_z\rangle$ is the most spectacular. 
One sees that on the average it does not get exactly to zero. 
 The zitterbewegung is thus exhibited quite clearly in these time
behaviour and qualitatively the component $\langle\sigma_z\rangle$ 
differs from the other two.
 
\subsection{Probability densities} \label{s4s2}
 We have not attempted to detect the zitterbewegung in the change of the
probability density with time. Indeed since this effect involves high
frequencies it is difficult to see in three dimensions. The reader is
invited to read \cite{nogami} % [7]
and \cite{toyama} % [8] 
where it has been shown in (1+1) dimensions.
 The counter-rotating wave that was discussed in section \ref{ss3} %3.3 
is however easily shown for high enough values of $r$. 
In Fig.~\ref{r5WP} %4 
the total probability density of the WP at the particular average radius 
(\ref{l19}) is represented in
spherical coordinates and for a few instants of time. What is shown was
entirely explained in \ref{ss3} % 3.3 
for each partial wave. A large part of the wave
stays at the initial position, essentially the part with spin up. The wave
is split into two waves which move in opposite direction and with the
same angular velocity. They are centered around a circle with $\theta=\pi/2$. 
Two analyses of the WP have then been made and are represented in 
Fig.~\ref{contpn} % 5 
and Fig.~\ref{contbisp} % 6 
for t=10. There one finds that both of these moving parts are mainly
localized in the second and third components of the spinor 
($|c2\rangle$ and $|c3\rangle$ respectively) 
and that the counter-rotating part is almost entirely
composed of negative energies. Here we are facing an effect totally absent
from a nonrelativistic behaviour and not understood in a one particle
theory. For lower values of $r$ this part of the wave is hardly visible (not
shown).
 
\subsection{FOLDY-WOUTHUYSEN representation}\label{s4s3}
  As derived in \cite{moreno} % [6] 
a FW transformation can be performed exactly on the DO. 
The result obtained is very simple and makes calculations extremely
easy. The small components $\Psi_2$ of the Dirac representation disappears and
eq.~(3a) results as the only equation valid for $\Psi_1$. The spinor has still
its eigenvalues given by (\ref{l5}) 
and the eq.~(\ref{l28}) and (\ref{l29}) should be used for
the spin averages. In other words the only relativistic effects are the use
of these energies. The effects introduced by negative energies
disappear. This fact was already well discussed in the (1+1) dimension model
\cite{toyama}.
A comparison of the calculation in the Dirac and the FW representation
enables to see exactly the manifestations of the zitterbewegung.
 Figures~\ref{r001fw} % 7 
and \ref{r5fw} % 8 are the same 
present in FW representation the same cases
as Figs. \ref{r001} % 1 
and \ref{r5} % 3 
in Dirac representation, respectively. Comparing the
figures one see indeed that the rapid fluctuations have been washed out. In
the FW representation the behaviour of each components of the spin is now
the same. Thus it is the use of the relativistic energies that produce now
the rapid but regular oscillations of the spin as well as the spread and
decay of its revival. 
 It is natural in this context to expect the disappearance of the
component rotating in the wrong sense. The WP are compared in 
Fig.~\ref{r5fwWP} % 9 
at time $t=10$. 
Only the part which rotates in the positive sense is left in the FW
representation.

\subsection{Other spin directions}\label{s4s4}
  The formulas of \ref{ss2} and \ref{ss3}  % 3.2 and 3.3 
can be combined conveniently to provide the
behaviour of a WP pointing initially in an arbitrary direction.
Such a study does not lead to a new dynamics. One can in this way 
simply put more weight on the part with spin down which is the most 
variable part. For example one can destroy almost totally the part not 
moving at the origin. Our choice of the initial direction has been made 
to see the components with spin up and down with the same magnitude. 
None of the other cases deserves a particular presentation. 

\section{Summary and conclusions}

  We have shown a new analogy between the relativistic DO and the JC model
of quantum optics. The time evolution of the average spin associated to a
WP in the DO is quite analogous to the time evolution of the occupation
numbers of each of a two levels atom which interacts with an
electromagnetic cavity. In the latter case the atom is entangled with the
cavity while in the former case the spin of the particle is entangled with
its orbital angular momentum according to rules fixed by Dirac equation. In
the same way in both models the mechanism of collapses and revivals takes
place. The collapse of the spin is compensated by a corresponding increase
of the orbital average angular momentum. This balance occurs periodically
in the nonrelativistic case \cite{arv94,arv95,rozm96}. % [9-11]. 
We have proposed the name of spin-orbit
pendulum for this effect. In the relativistic case the periodicity is
destroyed. There is then a rich behaviour of the spin components which are 
submitted to zitterbewegung. For a WP initially thrown with its velocity in
the $xOy$ plane the $z$ component of the spin contains more frequencies and
exhibits therefore most rapid oscillations.

    Related to the relativistic description we have found the presence of
a counter-rotating wave built mainly from negative energies states. This
component is particularly large for the geometry of the WP we have used in
our paper. Such a component is well identified in textbooks 
\cite{greiner} % [15] 
describing the scattering of a WP by a barrier where it is associated 
to Klein's paradox. 
To our knowledge it is the first time that such an effect has been
seen in the case of WP bound in a potential. We have been able to observe a
similar effect (however weaker) for a WP in 1+1 dimension. It is an open
question whether this effect also exists for a WP in a Coulomb field. The
smallness of the spin-orbit potential in this case may make observation 
difficult.
% does work for an easy observation. 

    Many relativistic effects are washed out in the FW representation. Our
conclusions confirm totally those of \cite{nogami,toyama}. % [7-8]. 
The coherence of the WP is also
lost because of the nonlinear relativistic energies. Therefore coherent
states of the harmonic oscillator generally spread. The counter-rotating
wave also disappears completely with negative energies and the dynamics
can be more interpreted with the ordinary one particle interpretation. 
This dynamics resembles then the well known dynamics of the population
inversion of the JC model with Eberly revivals of Rabi oscillations 
\cite{knight}. %  [12]. 
There is an attempt by Toyama and Nogami to provide coherent relativistic
WP of DO by using the inverse scattering method \cite{nogtoy}. % [16].
If these WP could be
defined also for a 3+1 oscillator we would have then probably a
relativistic spin orbit pendulum with a dynamics similar to the
nonrelativistic one. To our knowledge these WP have not yet been
constructed in the case where we take all dimensions into account.

\ackn
One of us (P.R.) kindly acknowledge support of Polish Committee for Scientific 
Research (KBN) under the grant 2 P03B 143 14.

\vfill

\Figures
\begin{figure}
% \hspace{10mm}
%  \resizebox{0.9\textwidth}{!}{%
%    \includegraphics{n20r.001dr.s1n.eps} }
\caption{Time evolution of average values of spin components for
$N=20$, $r=0.001$ in Dirac representation. Note that values of 
$\langle\sigma_z\rangle$ are 5 times enlarged.
The nonrelativistic case (dashed line) is included in the upper curve
for comparison (in this case the period is $T=2\pi/r=6283.185$).}
\label{r001} %1
\end{figure}

\begin{figure}
% \hspace{10mm}
%  \resizebox{0.9\textwidth}{!}{%
%    \includegraphics{n20r.025dr.s1.eps} }
\caption{The same as in Fig. \ref{r001} but for $r=0.025$.}
\label{r025} %2
\end{figure}

\begin{figure}
% \hspace{10mm}
%  \resizebox{0.9\textwidth}{!}{%
%    \includegraphics{n20r.5dr.s1.eps} }
\caption{The same as in Fig. \ref{r001} but for $r=0.5$.}
\label{r5} %3
\end{figure}

\begin{figure}
% \hspace{3mm}
%  \resizebox{0.95\textwidth}{!}{%
%    \includegraphics{n20r.5Dr.tfmul.nh.eps} }
\caption{WP motion for $N=20$, $r=0.5$ in Dirac representation.
Presented is the total probability density $|\Psi|^2$ on the surface of the
sphere with radius $r=x_0$. Note that motion of this {\em circular} WP
remains close to the equator (narrow $\theta$ range).}
\label{r5WP} %4
\end{figure}

\begin{figure}
% \hspace{10mm}
%  \resizebox{0.9\textwidth}{!}{%
%    \includegraphics{n20r.5Drpn.t10.nh.ps} }
% \vspace*{65mm}
\caption{Contributions from positive end negative energy states
 for $N=20$, $r=0.5$ $t=10$ in Dirac representation.}
\label{contpn} %5
\end{figure}

\begin{figure}
% \hspace{3mm}
%  \resizebox{0.95\textwidth}{!}{%
%    \includegraphics{n20r.5Dr.t10.tfmul.nh.ps} }
\caption{Contributions from all components of the bispinor $\Psi$
 (denoted as $c_1,c_2,c_3$ and $c_4$) 
 for $N=20$, $r=0.5$ $t=10$ in Dirac representation. Note 
 different vertical scales and fact that
 the contribution from the fourth component is zero.}
\label{contbisp} %6
\end{figure}

\begin{figure}
% \hspace{10mm}
%  \resizebox{0.9\textwidth}{!}{%
%    \includegraphics{n20r.001FW.s1.eps} }
\caption{Time evolution of average values of spin components for
$N=20$, $r=0.001$ in FW representation. Note that values of 
$\langle\sigma_z\rangle$ are 10 times enlarged.}
\label{r001fw} %7
\end{figure}
 
\begin{figure}
% \hspace{10mm}
%  \resizebox{0.9\textwidth}{!}{%
%    \includegraphics{n20r.5FW.s1.eps} }
\caption{The same as in Fig. \ref{r001fw} but for $r=0.5$.}
\label{r5fw} %8
\end{figure}
\begin{figure}
% \hspace{3mm}
%  \resizebox{0.95\textwidth}{!}{%
%    \includegraphics{n20r.5FW.tfmul.nh.ps} }
%\vspace{-10mm}
\caption{The same as in Fig. \ref{r5WP} but for FW representation.}
\label{r5fwWP} % 9
\end{figure}
\vfill*

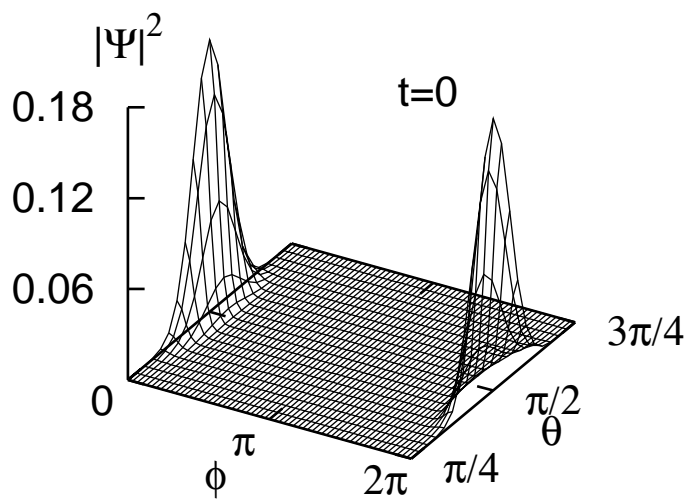
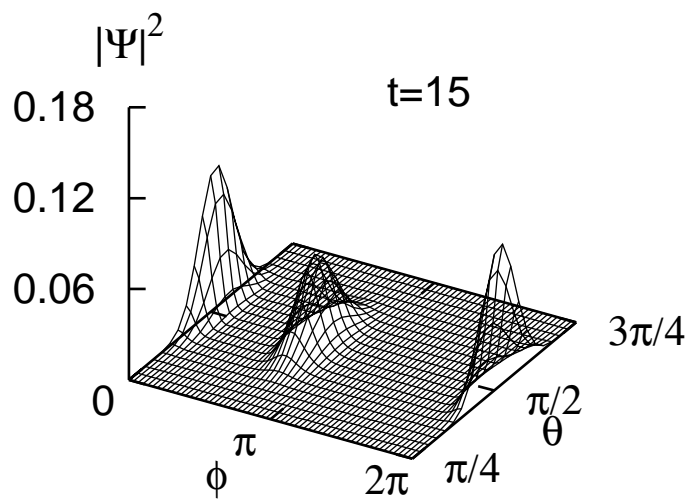
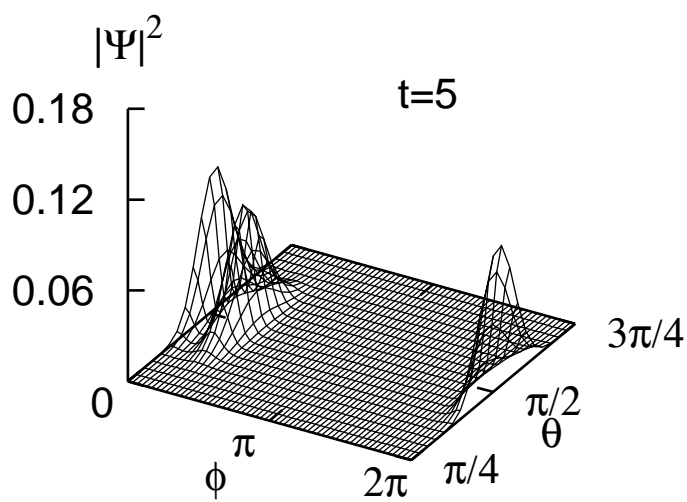
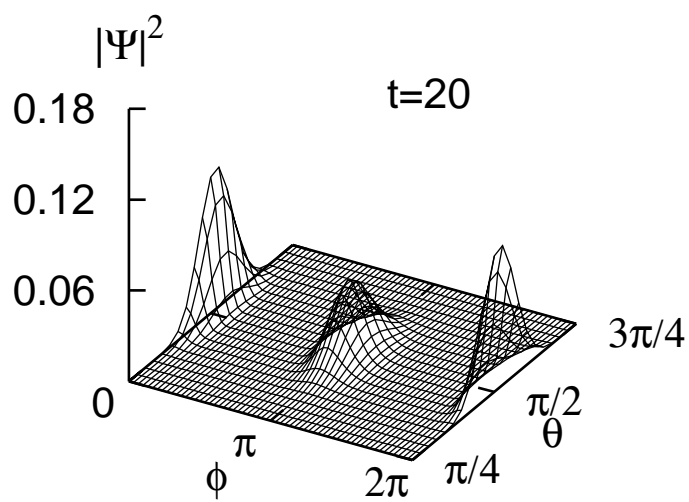
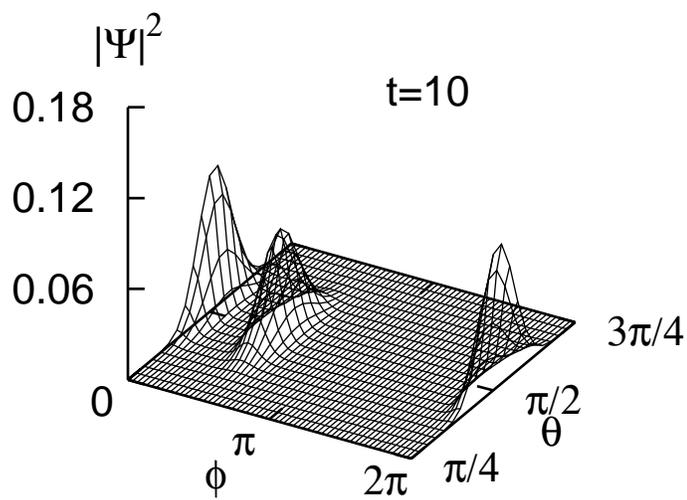
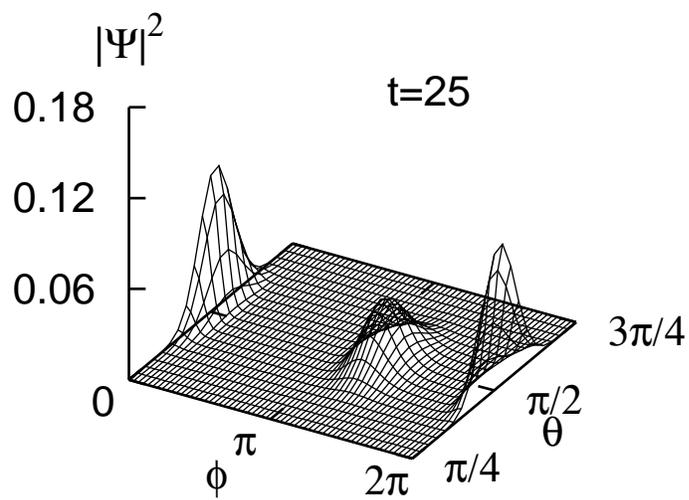
\end{document}